\documentstyle[12pt]{article}
\def\beqa{\begin{eqnarray}}
\def\eeqa{\end{eqnarray}}
\def\beq{\begin{equation}}
\def\eeq{\end{equation}}
\begin{document}
\def\bib#1{[{\ref{#1}}]}
\def\at{\tilde{a}}
\setcounter{page}{2}

\def\CcC{{\hbox{\tenrm C\kern-.45em{\vrule height.67em width0.08em depth-.04em
\hskip.45em }}}}
\def\RrR{{\hbox{\tenrm I\kern-.17em{R}}}}
\def\HhH{{\hbox{\tenrm {I\kern-.18em{H}}\kern-.18em{I}}}}
\def\DdD{{\hbox{\tenrm {I\kern-.18em{D}}\kern-.36em {\vrule height.62em 
width0.08em depth-.04em\hskip.36em}}}}
\def\ZzZ{{\hbox{\tenrm Z\kern-.31em{Z}}}}
\def\IiI{{\hbox{\tenrm I\kern-.19em{I}}}}
\def\NnN{{\hbox{\tenrm {I\kern-.18em{N}}\kern-.18em{I}}}}
\def\QqQ{{\hbox{\tenrm {{Q\kern-.54em{\vrule height.61em width0.05em 
depth-.04em}\hskip.54em}\kern-.34em{\vrule height.59em width0.05em depth-.04em}}
\hskip.34em}}}
\def\OoO{{\hbox{\tenrm {{O\kern-.54em{\vrule height.61em width0.05em 
depth-.04em}\hskip.54em}\kern-.34em{\vrule height.59em width0.05em depth-.04em}}
\hskip.34em}}}

\def\Csi{\Xi}
\def\id{{\rm id}}
\def\uno{{\bf 1}}
\def\uq2{U_q({\uit su}(2))}

\def\S{\Sigma}
\def\sh{{\rm sh}}
\def\ch{{\rm ch}}
\def\ad{A^{\dagger}}
\def\ac{a^{\dagger}}
\def\bd{B^{\dagger}}

\def\som#1{\sum\limits_{#1}}
\def\somma#1#2#3{\sum\limits_{#1=#2}^{#3}}
\def\integrale{\displaystyle\int\limits_{-\infty}^{+\infty}}
\def\intlims#1#2{\int\limits_#1^#2}
\def\tendepiu#1{\matrix{\phantom{x}\cr \longrightarrow\cr{#1\rightarrow +\infty
               \atop\phantom{x}}\cr}}
\def\tendezero#1{\matrix{\phantom{x}\cr \longrightarrow\cr{#1\rightarrow 0
               \atop\phantom{x}}\cr}}
\def\num{\strut\displaystyle}
\def\den{\over\displaystyle}
\def\fraz#1#2{{\strut\displaystyle #1\over\displaystyle #2}}
\def\bra#1{\langle~#1~|}
\def\ket#1{|~#1~\rangle}
\def\ave#1{\left\langle\left\langle~#1~\right\rangle\right\rangle}
\def\op#1{\hat#1}
\def\scalare#1#2{\langle~#1~|~#2~\rangle}
\def\esp#1{e^{\displaystyle#1}}
\def\qsp#1{q^{\displaystyle#1}}
\def\derpar#1#2{\fraz{\partial#1}{\partial#2}}
\def\part#1{\fraz{\partial}{\partial#1}}
\def\tg{{\rm tg}}

\def\ii#1{\item{$\phantom{1}#1~$}}
\def\ij{\item{$\phantom{11.}~$}}
\def\jj#1{\item{$#1.~$}}

\def\tens{\otimes}
\def\per{{\times}}
\def\co{\Delta}

\def\su2q{SU(2)_q}
\def\h1q{H(1)_q}
\def\edq{E(2)_q}
\def\etq{E(3)_q}
\def\so{SO(4)}
\def\soq{SO(4)_q}
\def\al#1{{\alpha}_{#1}}

\def\jt{J_3}
\def\jpm{J_{\pm}}
\def\jp{J_{+}}
\def\jm{J_{-}}
\def\ju{J_{1}}
\def\jd{J_{2}}
\def\j#1{J_{#1}}

\def\k#1{K_{#1}}

\def\nt{N_3}
\def\npm{N_{\pm}}
\def\np{N_{+}}
\def\nm{N_{-}}
\def\nu{N_{1}}
\def\nd{N_{2}}
\def\n#1{N_{#1}}

\def\pt{P_3}
\def\ppm{P_{\pm}}
\def\pp{P_{+}}
\def\pmen{P_{-}}
\def\pu{P_{1}}
\def\pd{P_{2}}
\def\p#1{P_{#1}}
\def\pol#1#2#3{{\cal P}_#1(#2,#3)}

\def\ee{\varepsilon}
\def\em{\ee^{-1}}

\def\e#1{{\bf e}_{#1}}

\def\epz#1{\esp{zJ_3^{#1}/2}}
\def\emz#1{\esp{-zJ_3^{#1}/2}}
\def\epw{\esp{wP_3/2}}
\def\emw{\esp{-wP_3/2}}
\def\epv{\esp{vM/2}}
\def\emv{\esp{-vM/2}}

\def\bin#1#2{\left[{\strut\displaystyle #1\atop\displaystyle #2}\right]_q}
\def\mapdown#1{\Big\| \rlap{$\vcenter{\hbox{$\scriptstyle#1$}}$}}
\def\mapcup{\bigcup}
\def\iinn{\rlap{{$\bigcup$}{$\!\!\! |$}}} 
\def\mapin{\iinn}

\begin{titlepage}
\title{Quantization of Damped Harmonic Oscillator,
\\ Thermal Field Theoris and q-Groups}

\author{Alfredo Iorio and Giuseppe Vitiello 
\\{\it Dipartimento di Fisica - Universit\`a di Salerno and INFN--Napoli}
\\{\it I84100 Salerno, Italy}
\\{E--mail: iorio@sa.infn.it vitiello@sa.infn.it}
\\
\\{\bf  Invited lecture at the Third International Workshop} 
\\{\bf on Thermal Field Theories, Banff, Canada, August 1993}}
              \date{\empty}
              \maketitle

\begin{abstract}
\noindent              
We study the canonical quantization of the damped harmonic oscillator by
resorting to the
realization of the q-deformation of the Wheyl-Heisenberg 
algebra (q-WH) in terms of finite difference operators. We relate 
the damped oscillator hamiltonian to the q-WH algebra and to the 
squeezing generator of coherent states theory. 
We also show that the q-WH algebra is the natural candidate
to study thermal field theory.
The well known
splitting, in the infinite volume limit, of 
the space of physical states into unitarily inequivalent 
representations of the canonical commutation 
relations is briefly commented upon in relation with the von Neumann 
theorem in quantum mechanics and with q-WH algebra. 
\end{abstract}

\thispagestyle{empty} \vspace{20. mm}

              \vfill
          \end{titlepage}
\noindent
In recent years much attention has been devoted to quantum 
deformations\cite{Bie},\cite{Mac} of 
Lie algebras in view of their  
great physical interest. It has been recognized\cite{Cel} that  
q-deformations appear whenever a discrete (space or time) length
characterizes the system under study.

\noindent
In this report we present recent results which relate 
the hamiltonian of the damped harmonic oscillator (dho)
to the q-WH algebra\cite{Ior} and to the 
squeezing generator for coherent states (CS)\cite{Cel2}. We also discuss the
relation of q-WH algebra with thermal field theory.

\noindent
In refs. 5-7 some aspects of dissipation in quantum field theory (QFT) 
have been studied by considering the canonical quantization of dho
\beq
m \ddot z + \gamma \dot z + \kappa z = 0 
\eeq
and it has been proven that
the space of the physical states splits into unitarily inequivalent
representations of the canonical commutation relations (ccr). Also, 
it has been realized that canonical quantization of the dho
leads to $SU(1,1)$ time-dependent CS, which are well known
in high energy physics as well as in quantum optics 
and thermal field theories. Moreover, dissipation phenomena 
and squeezed CS have been mathematically related, thus showing their
common physical features\cite{Cel2}.

\noindent
The hamiltonian describing an (infinite) collection of 
damped harmonic oscillators, is\cite{Cel3} 
${\cal H} = {\cal H}_{0} + {\cal H}_{I}$ where
\beq
{\cal H}_{0} = 
\sum_{\kappa} \hbar \Omega_{\kappa} {\Big (} A_{\kappa}^
{\dagger} A_{\kappa} - 
B_{\kappa}^{\dagger} B_{\kappa} {\big )}
\quad {\rm and} \quad 
{\cal H}_{I} = i 
\sum_{\kappa} \hbar \Gamma_{\kappa} {\big (} A_{\kappa}^{\dagger} 
B_{\kappa}^{\dagger} - A_{\kappa} B_{\kappa} {\big )} 
\eeq
where  ${\kappa}$ labels the field degrees of freedom, e.g. 
spatial momentum. As usual in QFT, we 
work at finite volume $V$ and 
perform at the end of the computations 
the limit $V \rightarrow \infty$. 
As well known, in order to set up the canonical formalism 
for a dissipative system, the doubling of the degrees 
of freedom is required; thus
the system $A$ is "doubled" by the system $B$ in Eq. (2).
The commutation relations are:
\beq
[\, A_{\kappa} , A_{\lambda}^{\dagger}\, ] = \delta_{\kappa , \lambda} = 
[\, B_{\kappa} , B_{\lambda}^{\dagger}\, ] \quad ; \quad [\, A_{\kappa} , 
B_{\lambda}^{\dagger}\, ] = 0 = [\, A_{\kappa} , B_{\lambda}\, ] 
\eeq
The group structure is
$\displaystyle{\bigotimes_{\kappa} SU(1,1)_{\kappa}}$.
We have $[\, {\cal H}_{0} , {\cal H}_{I}\, ] = 0$
and the ground state is (formally, at finite
volume $V$) 
\beq
|0(t)> = \prod_{\kappa} {1\over{\cosh{(\Gamma_{\kappa} t)}}} \exp{
\left ( \tanh {(\Gamma_{\kappa} t)} J_{+}^{(\kappa )} \right )} |0>
\eeq 
with $\displaystyle{J_{+}^{(\kappa )} \equiv 
A_{\kappa}^{\dagger} 
B_{\kappa}^{\dagger}}$.  Moreover, $
<0(t) | 0(t)> = 1~, \quad \forall t $~, and
\beq
<0(t) | 0> = \exp{\left ( - \sum_{\kappa} \ln \cosh 
{(\Gamma_{\kappa} t)~~} 
\right )}
\eeq
which shows how, provided ${\sum_{\kappa} 
\Gamma_{\kappa} > 0}$, 
\beq
\lim_{t\to \infty} <0(t) | 0> \, \propto \lim_{t\to \infty} 
\exp{( - t  
\sum_{\kappa}  \Gamma_{\kappa})} = 0 
\eeq  
Using~ the~ customary ~continuous ~limit relation ${ 
\sum_{\kappa} \mapsto {V\over{(2 \pi)^{3}}} \int \! d^{3}{\kappa}}$, 
in the infinite-volume limit we have (for ${\int \! d^{3} \kappa ~ 
\Gamma_{\kappa}}$ finite and positive)  
\beq
{<0(t) | 0> \rightarrow 0~~ {\rm as}~~ V\rightarrow \infty }
~~~\forall~  t  \quad , \quad
{<0(t) | 0(t') > \rightarrow 0~~ {\rm as}~~ V\rightarrow \infty}
~~~\forall~t~ {\rm and}~ t'~ ,~~~ t' \neq t
\eeq
Time-evolution transformations for creation and annihilation
operators are  
\beqa
A_{\kappa} \mapsto A_{\kappa}(t) &=& 
{\rm e}^{- i {t\over{\hbar}} {\cal H}_{I}} 
A_{\kappa} {\rm e}^{i {t\over{\hbar}} {\cal H}_{I}} =  
A_{\kappa} \cosh{(\Gamma_{\kappa} t)} - B_{\kappa}^{\dagger} \sinh{(
\Gamma_{\kappa} t)} ~ \nonumber \\
&& \\
B_{\kappa} \mapsto B_{\kappa}(t) &=& 
{\rm e}^{- i {t\over{\hbar}} {\cal H}_{I}} 
B_{\kappa} {\rm e}^{i {t\over{\hbar}} {\cal H}_{I}} =  
- A_{\kappa}^{\dagger} \sinh{(\Gamma_{\kappa} t)} + B_{\kappa} \cosh{(
\Gamma_{\kappa} t)} \nonumber 
\eeqa
and their hermitian conjugates. They can be implemented for every
$\kappa$, as inner automorphism for the algebra $su(1,1)_{\kappa}$. 
Such an automorphism is nothing but the well known 
Bogolubov transformations.
The transformations (8) are canonical, as 
they preserve the ccr (3).  In other words, at every time $t$ 
we have a copy
$\{ A_{\kappa}(t) , A_{\kappa}^{\dagger}(t) , B_{\kappa}(t) ,
B_{\kappa}^{\dagger}(t) \, ; \, | 0(t) >\, |\, \forall {\kappa} \}$ 
of the original algebra and of its highest weight vector $\{ A_{\kappa} ,
A_{\kappa}^{\dagger} , B_{\kappa} , B_{\kappa}^{\dagger} \, ; 
\, | 0 >\, |\, \forall {\kappa} \}$, induced by the time evolution operator 
(i.e we have a bona fide quantum realization of the operator algebra at 
each time t, which can be implemented by Gel'fand-Naimark-Segal construction 
in the C*-algebra formalism). The time evolution operator 
can therefore be thought of as a generator
of the group of automorphisms of ${\bigoplus_{\kappa}su(1,1)_{\kappa}}$ 
parametrized by time $t$.

\noindent
Let us point out that the various copies become unitarily inequivalent in the 
infinite-volume limit, as shown by Eqs.(7). 

\noindent
One can easily verify that   
$ A_{\kappa}(t) |0(t)> = 0 = B_{\kappa}(t) |0(t)> $ , for all~$t$.

\noindent
The commutativity of ${\cal H}_{0}$ with 
${\cal H}_{I}$ ensures that under time evolution the number $\,
{\left ( n_{A_{\kappa}} - n_{B_{\kappa}} \right )}\,$ is a constant of 
motion for any $\kappa$.    
The number of modes $A_{\kappa}$ is given, at each instant $t$, by 
\beq
{\cal N}_{A_{\kappa}} \equiv < 0(t) | 
A_{\kappa}^{\dagger} A_{\kappa} | 0(t) > 
= \sinh^{2}\bigl ( \Gamma_{\kappa} t \bigr ) 
\eeq
and similarly for the modes $B_{\kappa}$.  

\noindent
Moreover, 
\beqa
A_{\kappa}^{\dagger}(t) |0(t)> &=& {1\over{\cosh{(\Gamma_{\kappa} t)}}} 
A_{\kappa}^{\dagger} |0(t)>\, = 
{1\over{\sinh{(\Gamma_{\kappa} t)}}} B_{\kappa} 
|0(t)> \nonumber \\
&& \\
B_{\kappa}^{\dagger}(t) |0(t)> &=& {1\over{\cosh{(\Gamma_{\kappa} t)}}} 
B_{\kappa}^{\dagger} |0(t)>\, = 
{1\over{\sinh{(\Gamma_{\kappa} t)}}} A_{\kappa} 
|0(t)> \nonumber
\eeqa
Eq. (4) shows that $|0(t)>$ is a two-mode Glauber coherent state 
with equal numbers of modes $A_{\kappa}$ 
and $B_{\kappa}$ condensed in it for each $\kappa$ and each $t$.  
Eqs. (10) show that the creation of a mode 
$A_{\kappa}$ is equivalent to the destruction of a mode $B_{\kappa}$ and 
vice-versa.  This leads us to interpreting the $B_{\kappa}$ modes as the 
holes for the modes $A_{\kappa}$, and thus
the $B$-system can be considered as the sink where the 
energy dissipated by the $A$-system flows.

\noindent
We turn now our attention to the realization of q-WH algebra in terms
of finite difference operators\cite{Cel} in order to establish a formal 
relation with the dho hamiltonian.

\noindent
Since we want to preserve the analytic properties of
Lie algebra in the deformation procedure, we adopt
as a framework the Fock-Bargmann representation (FBR) 
in Quantum Mechanics (QM)\cite{Per} (we observe that, by working in the 
FBR, q-WH algebra is incorporated into the theory of the (entire) analytical 
functions, a result which is by itself interesting).

\noindent
The quantum version
of the WH algebra $[ a, a^\dagger ] = \IiI~, [ N, a ] = - a~, [ N,
a^\dagger ] = a^\dagger,$ is realized in terms of the operators $\{ a_q,
{\hat a}_q, N;  ~q \in {\cal C}\}$,  with relations
\beq
[ N, a_q ]= -a_q \;,\;  [ N, {\hat a}_q ] = {\hat a}_q \;,\;
[ a_q, {\hat a}_q ] \equiv a_q {\hat a}_q - {\hat a}_q a_q =
q^N 
\eeq
The FBR operators, solution of the WH commutation relations are
$N \to z {d\over dz}~, a^\dagger \to z~, a \to {d\over dz}~$.
The Hilbert space ~$\cal F$ is identified with the space 
of the entire analytic functions and
wave functions $\psi(z)$ are expressed as~
$ \psi (z) = \sum_{n=0}^\infty c_n u_n(z)~,
u_n(z) = {z^n\over \sqrt{n!}}~, (n~ \in {\cal I}_+)$. 
The set ~$\{ u_n(z)\}$~ provides an orthonormal basis in ~$\cal F$.
The realization of Eqs.(11) in the space ${\cal F}$ is obtained
as:
\beq
N \to ~z {d\over dz} , ~~~{\hat a}_q ~\to ~z~,
~~~~~a_q ~\to ~{\cal D}_q 
\eeq
with $z \in {\cal C}$ and the finite difference operator ${\cal D}_q$ 
defined by:
\beq
{\cal D}_q ~f(z) ~=~ {{f(q z) - f(z)}\over {(q-1) ~z}}~=~
{{q^{z {d\over {dz}}} - 1}\over{(q-1)~ z }}f(z)
\eeq 
with ~$f \in {\cal F}~, q = e^\zeta , ~\zeta \in {\cal C}$.
${\cal D}_q$ is the well known
q-derivative operator and, for $q \to 1$ (i.e.
$\zeta \to 0$), it reduces to the standard derivative. 
Note that ~${\hat a}_q = {\hat a}_{q=1} = a^\dagger$
~ and ~$\lim_{q\to1}
~a_q = ~a$. We can show that commutator  $[ a_q, {\hat a}_q]$ 
acts in ${\cal F}$ as follows\cite{Cel}
\beq
[ a_q, {\hat a}_q ]f(z)~=~q^{z{d\over dz}}f(z)~=~f(qz)~
\eeq
The q-deformation of the WH algebra is thus strictly related with
the finite difference operator ~${\cal D}_q$ ($q \not= 1$). 
This suggests to us that the q-deformation of the operator algebra 
should arise whenever we are in the presence of lattice or discrete
structure.
 
\noindent
Next, we show that the commutator $[ a_q, {\hat a}_q]$ acts as squeezing
generator, a result
which confirms the conjecture by which q-groups
are the  natural candidates to study the squeezed CS\cite{Cel5}.

\noindent
In the Hilbert space $~{\cal F}$ let us consider the harmonic oscillator 
hamiltonian
\beq 
H = {\hbar\omega} {\Big (} {\alpha}^\dagger{\alpha} + {1\over{2}}
{\Big )} 
\eeq
\beq
{\alpha} = {1\over {\sqrt{2\hbar\omega}}} \bigl({\sqrt{m}\omega} z + 
{i p_z\over{\sqrt{m}}}
\bigr) , ~~ { \alpha}^\dagger = {1\over {\sqrt{2\hbar\omega }}} 
\bigl({\sqrt{m}\omega} z -
{i p_z\over{\sqrt{m}}}\bigr), ~~~
[\alpha, \alpha^\dagger ] = \IiI
\eeq
where $z \in {\cal C}$, $p_z = - i \hbar {d\over {dz}}$ and $[ z, p_z]
= i \hbar$.
We have $2 z {d\over {dz}} f(z) = \bigl({\alpha}^2 -
{\alpha}^{\dagger 2}\bigr) f(z) - f(z), ~~f \in {\cal F} $~, and 
on any state $\psi (z)$
\beqa
[ a_q, {\hat a}_q] \psi (z)  &=&
  \exp{\Big (}\zeta z {d\over dz}{\Big )}  \psi (z) ~=~ 
{1\over{\sqrt q}} \exp{\Big (}{\zeta\over 2}{\big (}\alpha^2 - 
{\alpha^\dagger}^2{\big )} {\Big )}  \psi (z)  \nonumber \\
&& = {1\over{\sqrt q}}{\hat {\cal S}}(\zeta) \psi (z) = 
{1\over{\sqrt q}} \psi_{s}(z)
\eeqa
with $q = e^\zeta$, $\psi _s(z)$ denoting the squeezed state and
${\hat {\cal S}}(\zeta)$ the squeezing generator.

\noindent
We thus conclude that $\zeta = \log~q$
plays the r\^ole of the squeezing parameter
and the commutator $[ a_q, {\hat a}_q ]$ 
is, up to a factor, the squeezing generator with respect 
to the $\alpha$ and $\alpha^\dagger$ 
operators:
\beq
{\sqrt q}[ a_q, {\hat a}_q] ~=~{\hat {\cal S}}(\zeta)~=~
\exp{\bigg (}{\zeta \over 2}\bigl(\alpha^2 -
{\alpha^\dagger}^2\bigr){\bigg )}
\eeq
On the other hand we also observe that
the right hand side of (18) is an $SU(1,1)$ group element. In fact by
defining ${1\over  2}\alpha ^2 = K_{-}$,
${1\over  2}\alpha^{\dagger 2} = K_{+}$, 
${1\over 2}(\alpha^\dagger 
\alpha + {1\over 2})
= K_{z} = {1\over{2 \hbar \omega}} H$, 
we easily check they close the $su(1,1)$ algebra.

\noindent
Let us now observe that the operator 
${\sqrt q}[ a_q, {\hat a}_q]~=~{\hat {\cal S}}(\zeta) $
generates the Bogolubov transformations
\beq
\alpha(\zeta) = {\hat{\cal S}}(\zeta)^{-1}
\alpha {\hat {\cal S}}(\zeta) =
\alpha \cosh{\zeta} - \alpha^{\dagger} \sinh{
\zeta}
\eeq
and h.c. By inverting Eqs.(16) and by using Eq.(19) we obtain
\beq
z(\zeta) = z_{0}\exp (-{\zeta})
\eeq
By setting
$\zeta = {\gamma\over{2m}} t \equiv \Gamma t $,
from the squeezed $z(\zeta)$, Eq.(20), we obtain the damped amplitude of the 
classical dho with motion equation (1). 

\noindent
Summarizing, in quantizing the dho, Eq.(1),  one 
may proceede as follows: start
by considering the classical
harmonic oscillator equation
\beq
m \ddot z + \kappa z = 0
\eeq
Quantize it in the standard way by introducing the 
creation and annihilation operators (cf. Eqs.(16)).
Next, introduce the q-deformation of the WH algebra in FBR and the 
associated ${\hat {\cal S}}(\zeta)$ operator as in Eq.(18), 
with 
$\zeta = {\gamma\over{2m}} t \equiv \Gamma t $~. 
${\hat {\cal S}}(\zeta)$ is thus the time evolution
generator of the quantized dho
$\alpha(t)$~(cf. the Bogolubov transformations (19), see also Eqs.(8)). 
Finally, introduce the 
shifted frequency ${\Omega}^{2} = {{\kappa} \over{m}} - {{\gamma^{2}}
\over {4m^{2}}}$ for the damped solution.

\noindent
The finite life-time $\tau = {1 \over{\Gamma}}$, which characterizes the 
damping, acts as discrete time-length for the system 
(time-energy uncertainty relation
implies $\Delta t \geq {1\over{\Gamma}}$) and thus
requires the
use of the q-WH algebra 
according to the above conclusions on the occurrence of 
q-deformations in the presence of a finite (time) length. 

\noindent
To see the relation with the QFT hamiltonian, eq.(2), we label with
$\kappa$ the operators $\alpha$ and $\alpha ^{\dagger}$,
the decay constant $\Gamma$ and other appropriate quantities
and introduce convenient summations over $\kappa$ in the exponentials.
Then, we introduce the "doubled" degree of freedom $\beta_{\kappa}$ 
(commuting with $\alpha_{\kappa}$).
The corresponding q-WH algebra for $b_{\kappa,q'}$ operators 
(commuting with $a_{\kappa,q}$) is
introduced with $q' = \exp(-\zeta_{\kappa})$. Next, we consider
the double mode squeezing operator
\beq
[ a_{\kappa,q}, {\hat a}_{\kappa,q}][ b_{\kappa,q'}, 
{\hat b}_{\kappa,q'}] =
\exp{\big (}{{\zeta_{\kappa}} \over 2}{\big [}{\big (}{\alpha_{\kappa}} ^2 -
{\alpha_{\kappa}} ^{\dagger 2} {\big )} - {\big (}{\beta_{\kappa}}^2 -
{\beta_{\kappa}}^{\dagger 2} {\big )}{\big ]} {\big )}
\eeq
We observe that the right hand side of Eq.(22) is an $SU(1,1)$ group
element. In fact, by using the linear canonical transformations
\beq
{A_{\kappa}}~=~{({1\over{\sqrt{2}}})(\alpha_{\kappa} +
\beta_{\kappa})}, ~~~~~{B_{\kappa}}~=~
{({1\over{\sqrt{2}}})(\alpha_{\kappa} -  \beta_{\kappa}) ~}
\eeq
we have $[({\alpha_{\kappa}} ^2 -
{\alpha_{\kappa}} ^{\dagger 2} ) - ({\beta_{\kappa}}^2 -
{\beta_{\kappa}}^{\dagger 2} )] = - ({A_{\kappa}} ^{\dagger}
{B_{\kappa}} ^{\dagger} - A_{\kappa}B_{\kappa}) =
- (J_{+} - J_{-})~$, which together with 
$J_{3} = {1\over{2}}({A_{\kappa}} ^{\dagger}A_{\kappa} +
{B_{\kappa}} ^{\dagger}B_{\kappa} + 1)$
close the $su(1,1)$ algebra. 

\noindent
By setting ${\zeta}_{\kappa} = {\Gamma}_{\kappa} t$, Eq. (22) leads to 
nothing but the time
evolution operator associated to the hamiltonian ${\cal H}_{I}$
(cf.eq.(2)):
\beq
\prod_{\kappa}{[ a_{\kappa,q}, {\hat a}_{\kappa,q}][ b_{\kappa,q'}, 
{\hat b}_{\kappa,q'}]} ~=~ 
\exp{\big (}{i\over{\hbar}} {\cal H}_{I}t {\big )}
\eeq
thus recovering the QFT formalism of quantum dho\cite{Cel3}.

\noindent
We conclude that q-deformation turns out to be a powerful tool in
the discussion of dissipative phenomena in QFT. 
We note that inequivalent representations $|O(t)>$ 
are parametrized by the q-deformation parameter through the time
dependence of ${\zeta(t)}$. 
By recalling that the representation ${|O(t)>}$ for the dho has been
shown\cite{Cel2}-\cite{Cel4} to be the same as the 
TFD\cite{Tak} representation ${|O(\beta (t))>}$,
we also recognize the strict relation between q-WH algebra and 
finite temperature QFT. As a matter of fact, even independently of
the above discussion of the dho quantization, the derivation of 
Eq.(22) holds in full generality and by setting $\zeta_{\kappa} =
\theta_{\kappa}(\beta)$ we see that the operator
$[ a_{\kappa,q}, {\hat a}_{\kappa,q}][ b_{\kappa,q'}, 
{\hat b}_{\kappa,q'}]$ 
is indeed the generator of the Bogolubov transformations in
conventional TFD\cite{Tak}: q-WH algebra thus appears as the natural
candidate to study thermal field theories.

\noindent
A further relation which can be established\cite{Vit}  
is the one between q-WH algebra and the canonical 
quantization in curved space-time background, e.g. in the
presence of classical gravitational background\cite{Mar}. It is
interesting to note that in such a case the q-deformation can be 
thought as  induced by the gravitational background.

\noindent
The above remarks strongly suggest to us that q-deformations
are mathematical structures which are charateristic of basic, deep
features of QFT. As a preliminary investigation in this direction
we have studied
the relation
of q-deformation with the von Neumann theorem in QM:
we can show\cite{Ior2} that in the infinite volume limit
each one of the unitarily inequivalent representations of the ccr
is associated with a given value of the 
q-deformation parameter of the q-WH algebra.

\end{document}